# FQA: A Full-Space Quantization-Driven Architecture for Hardware-Efficient Piecewise Approximation of Nonlinear Activation Functions

Chenjun Hao, Feng Yan, Hongbing Pan , and Yuxuan Wang

*Abstract*—**In this paper, we propose a full-space quantization-driven architecture (FQA) for the hardware-efficient piecewise polynomial approximations (PPAs) of nonlinear activation functions. FQA comprehensively considers both fractional-bit truncation error and quantization error that cause the deviation of the optimal approximation coefficients. Crucially, FQA can precisely determine and search the complete range of optimal coefficients. Based on the proposed FQA, we develop two distinct hardware implementation schemes to cater to different resource-performance trade-offs. Furthermore, we decouple all the fractional word lengths (FWLs) involved in the calculation process to enable the exploration of superior hardware architectures. To mitigate the increased software computation time caused by the expanded quantization space, we design an acceleration method named TBW (target-guided bisection window) to expedite the piecewise calculation and searching process. Experimental results demonstrate that, compared to existing architectures, FQA can significantly reduce the number of required segments while achieving the optimal Maximum Absolute Error (MAE). For the hardware design of the Sigmoid function, our approach achieves over 50% reduction in area and power consumption compared to the state-of-the-art PPA architecture. Finally, we present a complete design workflow for deploying PPA on configurable hardware, maximizing the utilization of existing hardware resources and minimizing MAE.**

## I. INTRODUCTION

Driven by the rapid progress of artificial intelligence, the hardware-efficient realization of nonlinear activation functions (NAFs) has become a key enabler for energy-efficient and high-performance neural accelerators. In deep neural networks (DNNs), NAFs such as the Sigmoid and Tanh functions are critical for enabling both deep learning capabilities and the nonlinear expressiveness of neural networks, and their hardware implementations can directly affect the performance of neural network accelerators [1], [2]. Recently, Kolmogorov-Arnold Networks (KANs) have introduced learnable activation functions as a central architectural element [3]. Unlike conventional models, the activation functions in KANs evolve during training into non-standard forms, significantly escalating the complexity of hardware design. These NAFs cannot be directly decomposed into linear algebraic combinations and must be transformed into hardware-implementable structures through appropriate approximation techniques.

In the approximation of NAFs, implementation methods can be broadly categorized into iterative and non-iterative approaches, depending on whether the computation involves loop-based iterations. While iterative methods can improve approximation accuracy by increasing the number of iterations, this inevitably leads to longer latency and higher computational overhead [4], [5], [6], [7]. Compared to iterative approaches, non-iterative methods are favored for their superior computational speed. Common non-iterative techniques include table-addition methods [8], [9], [10], [11], [12], [13]and piecewise polynomial approximation (PPA) methods[14], [15], [16], [17], [18], [19], [20], [21], [22], [23], [24], [25], [26], [27], [28], [29], [30], [31], [33], [34], [35]. Owing to their reliance solely on adders, table-based addition methods typically offer low circuit latency. However, the size of the LUT grows exponentially with the input word lengths (WLs), making these methods practical only for scenarios with relatively low input resolution [7].

The PPA method typically partitions the target approximation interval into multiple subintervals, with each subinterval fitted by a polynomial function. Based on whether the segment lengths are uniform, PPA methods can be further classified into uniform segmentation approaches and non-uniform segmentation approaches [20], [21], [22], [23], [24], [25], [26], [27], [28], [29], [30], [31], [33], [34], [35]. MBS, TEA-S, and TEA-SPS focus on the customized hardware architectures for Softmax functions [33], [34], [35]. When high approximation accuracy is required, non-uniform segmentation methods offer significant area advantages over uniform segmentation [30]. Within each subinterval, the PPA approach determines polynomial coefficients through dedicated fitting algorithms. Commonly used fitting techniques include Taylor series expansion[16], [17], the Chebyshev method[30], the minimax algorithm[15], [16], [17], [18], [22] and the Remez exchange algorithm [31], [32]. QPA contends that the Remez algorithm exhibits superior performance over alternative fitting methods in terms of approximation accuracy [31]. Regarding the evaluation criteria for PPA, references [19], [20] utilize relative error as the minimum upper bound of approximation error to determine the segmentation strategy for function partitioning. However, when hardware efficiency is prioritized, absolute error is generally considered a more appropriate metric [21]. Consequently, most recent studies adopt the maximum absolute error (MAE) as the primary evaluation standard [25], [26], [27], [28], [29], [30], [31], [33], [34], [35].

To implement the results of the PPA method in hardware, it is necessary to quantize both the polynomial coefficients and the intermediate computational steps. In [25], the issue of quantization was not addressed prior to hardware realization. PLAC introduces a software-based quantizer. However, it encounters significant challenges in determining the optimal QF values.[26]. While SQ performs an intercept readjustment after slope quantization, this method is specifically designed for the approximation of floating-point logarithmic operators [28]. ML-PLAC extends the SQ-based quantization strategy to other nonlinear functions [29]. QPA introduces a fine-tuning scheme during coefficient quantization, resulting in a lower MAE compared to previous works [31].

The WL allocation represents another key issue in the quantization phase. PLAC binds the fractional word lengths (FWLs) of the slope k, intercept b, multipliers, and adders to a single unified parameter *qw*, which may lead to precision waste and unnecessary hardware overhead [26]. In ML-PLAC, to effectively replace multipliers with shift-and-add units, the FWLs of slope k is typically constrained to a small value[29]. QPA adopts a more flexible FWLs configuration for arithmetic units, thereby reducing hardware area. However, it still binds the FWLs of the multiplier output to that of the adder, leaving potential hardware optimization opportunities unexploited [31].

In addition, with respect to segmentation strategies, [25] adopts a sequential approach that starts from the end of each interval and incrementally searches for the first point satisfying the MAE constraint, resulting in relatively slow segmentation. To accelerate this process, [26] introduces a bisection method; however, segmentation and quantization are performed separately, leading to limited overall efficiency. In [28], the quantization process is integrated into the binary method, significantly improving segmentation efficiency. This integrated approach has been widely adopted in subsequent works such as [28], [29], [31].

In this work, we propose full-space quantization-driven architecture (FQA), which incorporates a dedicated quantization method designed to systematically explore the full coefficient space for hardware-efficient PPA. At the same time, to accelerate the computation process, we propose a target-guided bisection window method (TBW) for efficient hardware design of NAFs with controllable maximum absolute error. The proposed approach not only achieves higher accuracy and efficiency compared to other methods, but also maintains universality across all types of NAFs.

The main contributions of this work are as follows:

1) We propose FQA, a hardware-efficient architecture powered by a dedicated quantization algorithm that addresses the expansion of the search space arising from the interplay between truncation and quantization errors. By exhaustively exploring the coefficient quantization space, the FQA-based quantization method identifies the optimal coefficient set that yields the minimum MAE. Based on the proposed FQA, we develop two distinct hardware implementation schemes: FQA-On and FQA-Sm-On.

2) We propose TBW, which accelerates the segmentation process and reduces computational cost by leveraging a pre-estimated target segment range. It is worth noting that TBW is independent of the FQA method and can be applied to other quantization approaches where the target number of segments is predetermined.

3) We refine the hardware optimization strategy by introducing techniques such as flexible coefficient WL configuration. Specifically, we decouple the FWLs of all arithmetic components and adopt a FWL stitching strategy between components with different fractional precisions, enabling more flexible coefficient representation and further reducing hardware area.

4) Considering the requirement for coefficient configurability to achieve high-flexibility activation functions, a hardware-constrained PPA workflow is developed based on FQA. The proposed framework maximizes the utility of available hardware resources to achieve the highest possible approximation precision under specific implementation constraints.

The remainder of this paper is organized as follows. Section II introduces the fundamental principles and the state-of-the-art of the PPA architecture. Section III details the theory behind the FQA and the TBW method for accelerating piecewise operations, followed by the introduction of the FQA-On and FQA-Sm-On implementation schemes. The hardware-constrained PPA workflow is also introduced in this section. Section IV provides the ASIC implementation results of the proposed work and offers a comparison with previous studies. Finally, Section V presents the conclusion of this paper.

## II. OVERVIEW OF THE PPA METHOD

In this section, we present the basic principles of a general PPA methodology. Section II-A describes the algorithm design phase of the PPA process, while Section II-B focuses on the hardware design phase of the PPA process. In Section II-C, we analyze the advantages and limitations of the state-of-the-art quantization-aware PPA architecture (QPA) proposed in [31].

### A. Algorithm Design Phase of the PPA Process

In existing works, typical PPA methods generally consist of four components: fitting, quantization, segmentation, and hardware design. Among these, the first three are implemented on the software side. In the PPA design flow, the NAF $f(x)$ is partitioned into multiple segments over the input domain. In the $i$-th segment, the corresponding approximation polynomial is denoted as $h_i(x)$, which is expressed as

$$h_i(x) = (...(a_{i,1}x + a_{i,2})x + ... + a_{i,n})x + b_i \qquad (1)$$

where $\{a_{i,0}, a_{i,1}, \ldots, a_{i,n}, b_i\}$ are the coefficients for the $i$-th segment.

The structure of the polynomial $h_i(x)$ is illustrated as the polynomial computing unit in Fig. 1. The MAE is a commonly used metric in PPA methods[25], [26], [27], [28], [29], [30], [31]. Its expression prior to quantization is given by:

$$MAE_{soft} = \max(|f(x) - h(x)|) \tag{2}$$

To ensure consistency between software results and hardware computation results, the polynomial $h(x)$ is typically quantized into $h_q(x)$, and the corresponding hardware-side maximum absolute error ($MAE_{hard}$) is computed. Its expression is given by:

$$MAE_{hard} = \max(|f(x) - h_q(x)|) \tag{3}$$

The objective of the fitting algorithm is to determine the coefficients of the polynomial that minimize $MAE_{soft}$. Among existing methods, the Remez algorithm[31] is capable of achieving the lowest $MAE_{soft}$ through iterative optimization. However, in practical applications, the coefficients often require quantization to a specific WL. When conventional quantization methods are applied to the coefficients obtained by the Remez algorithm, they introduce additional errors, consequently resulting in a large $MAE_{hard}$. Consequently, the optimal coefficient values in this quantized domain may significantly deviate from the optimal coefficient values before quantization.

During the quantization process, it is first necessary to determine the WL for each computational stage and then calculate the final coefficient values using a quantization algorithm. The quantization algorithm is a critical factor affecting $MAE_{hard}$. Traditional methods apply fixed quantization rules such as ceil, floor, or round at each quantization step [26], [29], [30]. QPA introduces a ±1 fine-tuning strategy, which effectively expands the quantization space to include all three rounding modes, often resulting in a lower $MAE_{hard}$.

However, QPA does not specify an upper bound for the fine-tuning range. The ±1 fine-tuning strategy alone is insufficient to achieve higher computational accuracy, while unbounded expansion of the fine-tuning range would incur excessive computational overhead.

After determining the fitting and quantization algorithms, it is also necessary to select an appropriate segmentation algorithm. In current PPA methods, a widely used segmentation strategy is the bisection method proposed by Dong et al. [26], which can efficiently identify the minimum number of segments with relatively low computational cost. Compared with the strategy introduced by Sun et al. [25], this method significantly reduces computation time. However, when the number of target segments becomes large, the bisection method still incurs substantial redundant computation.

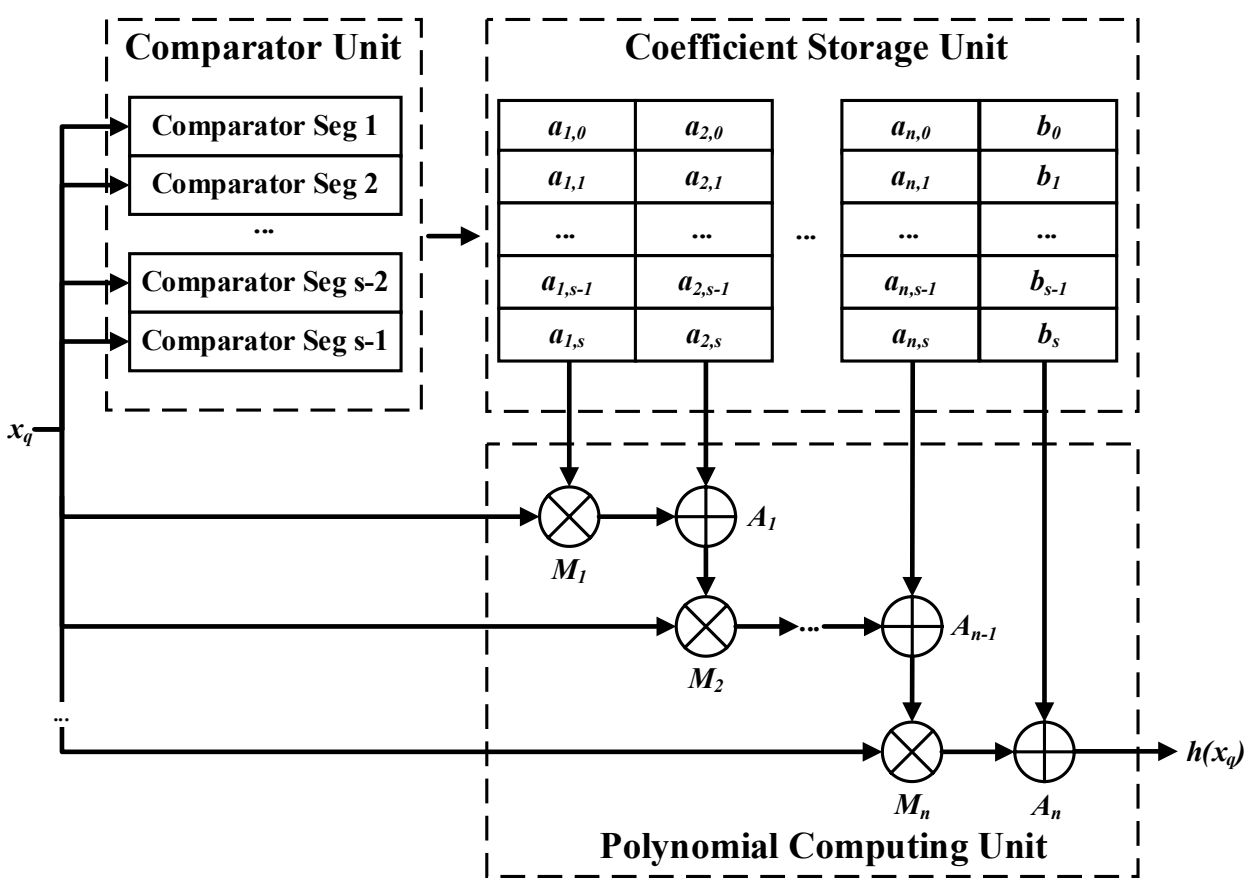

Fig. 1 Typical PPA hardware architecture

### B. Hardware Design Phase of the PPA Process

The hardware architecture adopted by most recent PPA implementations is illustrated in Fig. 1 and has been widely used in prior works [25], [26], [27], [28], [29], [30], [31]. This architecture primarily consists of three components: an index generator, a parameter memory, and a computation unit. The index generator is typically composed of multiple comparators. Assuming the number of segments is $s$, the design usually requires $s$-1 comparators to determine the segment index corresponding to the input $x$. The retrieved index is then used to access the corresponding polynomial coefficients and from the parameter memory, which are subsequently fed into the computation unit for evaluation.

The $n$-order computation unit typically consists of $n$ multipliers and $n$ adders. In ML-PLAC [29], the multipliers are replaced with shifters and adders, significantly reducing hardware area. However, this replacement leads to a larger $MAE_{hard}$, limiting computational accuracy and making it applicable only in specific scenarios. Clearly, the FWLs of individual components within the computation unit have a substantial impact on the overall hardware area. Therefore, it is critical to establish a design flow that balances computational accuracy and hardware efficiency.

### C. Advantages and Disadvantages Analyses

QPA, recognized as one of the most advanced existing PPA architectures, integrates several strengths from prior works. The key advantages of this approach are as follows:

1. QPA adopts the Remez approximation algorithm, which improves approximation accuracy through iterative refinement.

2. To reduce quantization error, QPA introduce a fine-tuning strategy during coefficient quantization, which enlarges the search space beyond conventional methods.

3. QPA further reduce hardware cost by decoupling the FWLs of partial polynomial coefficients.

Despite these advantages, QPA still presents several limitations that restrict its effectiveness in high-precision and resource-constrained scenarios, as detailed below:

1. QPA adopts the Remez algorithm to obtain high-

precision polynomial coefficients prior to quantization. However, due to quantization-induced distortion, the optimal coefficients after quantization may deviate significantly from the original unquantized results. As a result, the iterative refinement performed by Remez becomes redundant and wasteful, introducing excessive computational overhead without corresponding gains in final hardware accuracy.

2. The fine-tuning strategy used in QPA expands the search space compared to basic floor, ceil, or round operations, but it remains insufficient for cases demanding high accuracy or tighter $MAE_{hard}$ bounds. Moreover, QPA does not define an upper bound for the search space, potentially missing better quantized solutions located beyond the ±1 range.

3. QPA employs the bisection method from PLAC [26], which reduced computation compared to [25], but still incurred redundant evaluations, especially when large numbers of segments were required.

4. QPA maintains a fixed binding between the FWLs of multiplier outputs and adder outputs. This rigid configuration limited the flexibility of hardware design and resulted in excessive hardware area consumption.

## III. PROPOSED APPROXIMATION METHODOLOGY

In this section, we progressively introduce FQA and the TBW method developed to accelerate the computation speed. We first introduce the overview of FQA in Section III-A, followed by a detailed description of the TBW method in Section III-B.

Subsequently, two hardware implementation schemes, namely FQA-On and FQA-Sm-On, are presented in Section III-C and Section III-D, respectively. FQA-On targets the classical PPA hardware architecture, where $n$ denotes the polynomial order. FQA-Sm-On is designed to replace the first stage multipliers with shift-add units, aiming to reduce hardware area, with n representing the number of shifters used. Finally, Section III-E proposes a hardware-constrained PPA workflow.

### *A. Overview of FQA*

We first consider the determination of the FWL. In the hardware design of PPA, the FWL plays a critical role in determining both hardware area and computational accuracy. Typically, the FWLs of the input and output are specified by application-level requirements. The key design concern lies in configuring the FWLs of the internal multipliers and adders. We first analyze the FWL strategy for multipliers. Considering that multipliers occupy a significant portion of the hardware area, we face two main considerations. On the one hand, to effectively reduce the area of the multipliers and other components, the FWL of the output is typically set to be less than the sum of the two input FWLs [26], [29], [30], [31]. However, this strategy introduces a truncation error, which causes a deviation from the optimal coefficients. On the other hand, the FWL of the multiplier output must not be excessively small. Otherwise, it may lead to underutilization of coefficient precision and a significant increase in the number of segments. This not only degrades approximation accuracy but also fails to reduce overall hardware cost effectively. Therefore, the multiplier output FWL is typically chosen between the input FWL and the full-precision FWL (i.e., the sum of the input and coefficient FWL without truncation).

Next, we consider the FWL allocation strategy for the addition coefficients in the stage following the multipliers. The FWL of the addition coefficient does not necessarily have to be equal to that of the preceding multiplier's output. The design accounts for potential FWL mismatches:

Case 1: $W_{add} < W_{mult,out}$

If the FWL of the addition coefficient is smaller, the superfluous fractional bits from the multiplier output are preserved. These preserved bits are then concatenated with the adder's output to yield the final computation result.

Case 2: $W_{add} \geqslant W_{mult,out}$

In certain scenarios, assigning a larger FWL to the addition coefficient than that of the preceding multiplier output can surprisingly result in better hardware efficiency. In such a configuration, the adder input FWL is set equal to the multiplier output FWL ($W_{adder} = W_{mult,out}$). The excess fractional bits of the addition coefficient are then preserved and concatenated with the adder output to form the final result. This setup serves as a complementary option within the design space and can be selectively explored for further optimization.

Crucially, the addition stage is intended to finely adjust the result from the previous stage. Therefore, its FWL must not be overly reduced. An insufficient FWL would cause the stage to lose its ability to properly compensate, effectively wasting the precision gained in the preceding multiplier operation. Although the FWL of the addition coefficient has a limited degree of hardware optimization, this part of the optimization space cannot be ignored, especially the high-order polynomial approximation method.

Fig. 2 illustrates one of the detailed polynomial computation circuits utilized by our FQA. Here, $W_i$ denotes the FWL of the input $x_q$, $W_{a,1}$ to $W_{a,n}$ and $W_b$ denote the FWLs of the polynomial coefficients, and $W_{o,1}$ to $W_{o,n}$ represent the FWLs of multiplier outputs. $W_{m,1}$ to $W_{m,n}$ denote the FWLs of the inputs for the second up to the $n$-th multipliers, where $W_{m,n}$ is max($W_{a,n}$, $W_{o,n}$).

To reduce adder resource, a concatenation-based structure is adopted for the adders, as shown in Fig. 3. In Fig. 3(a), the coefficient used in the addition stage has a smaller FWL, whereas in Fig. 3(b), the preceding multiplier output has a smaller FWL. In both structures, the FWL of the adder is set to the minimum value between the FWL of the preceding multiplier output and the corresponding addition coefficient (i.e., min ($W_{mult,out}$, $W_{add}$)). The superfluous fractional bits are excluded from the adder calculation. Instead, they are concatenated with the adder's computation result after the operation is complete to form the final output of the current calculation. FQA fully decouples the FWLs of all arithmetic units, greatly expanding the design space for PPA.

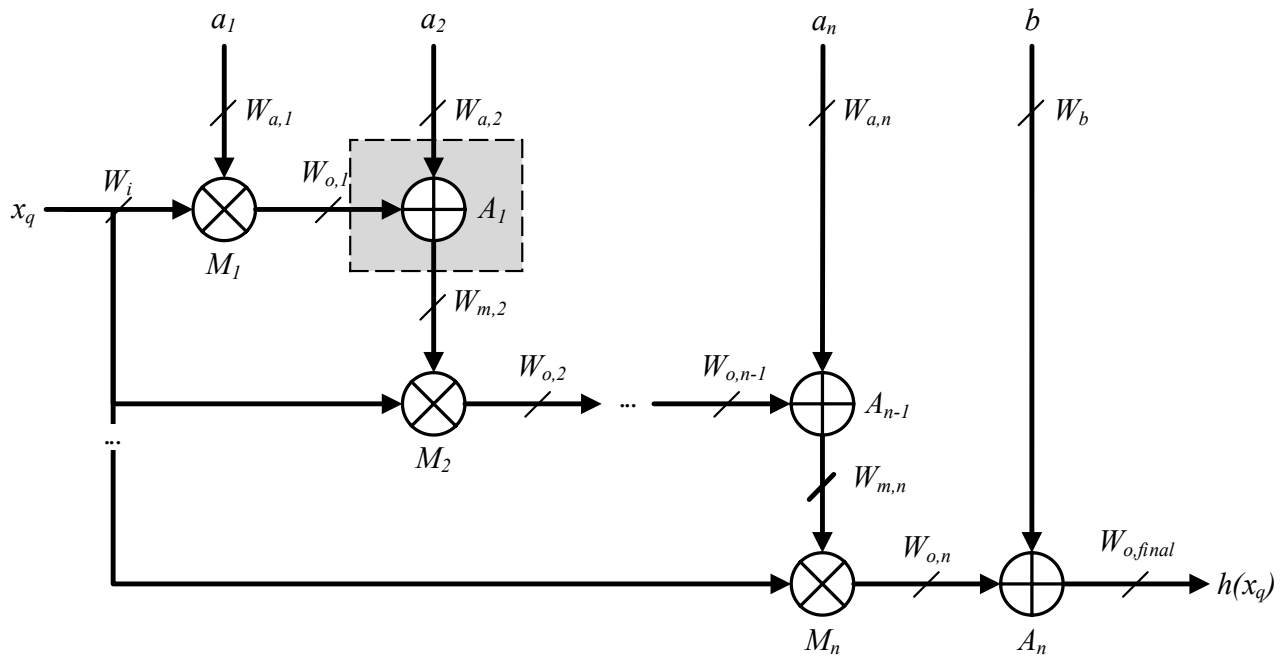

Fig. 2 Calculation circuit corresponding to FQA-On scheme

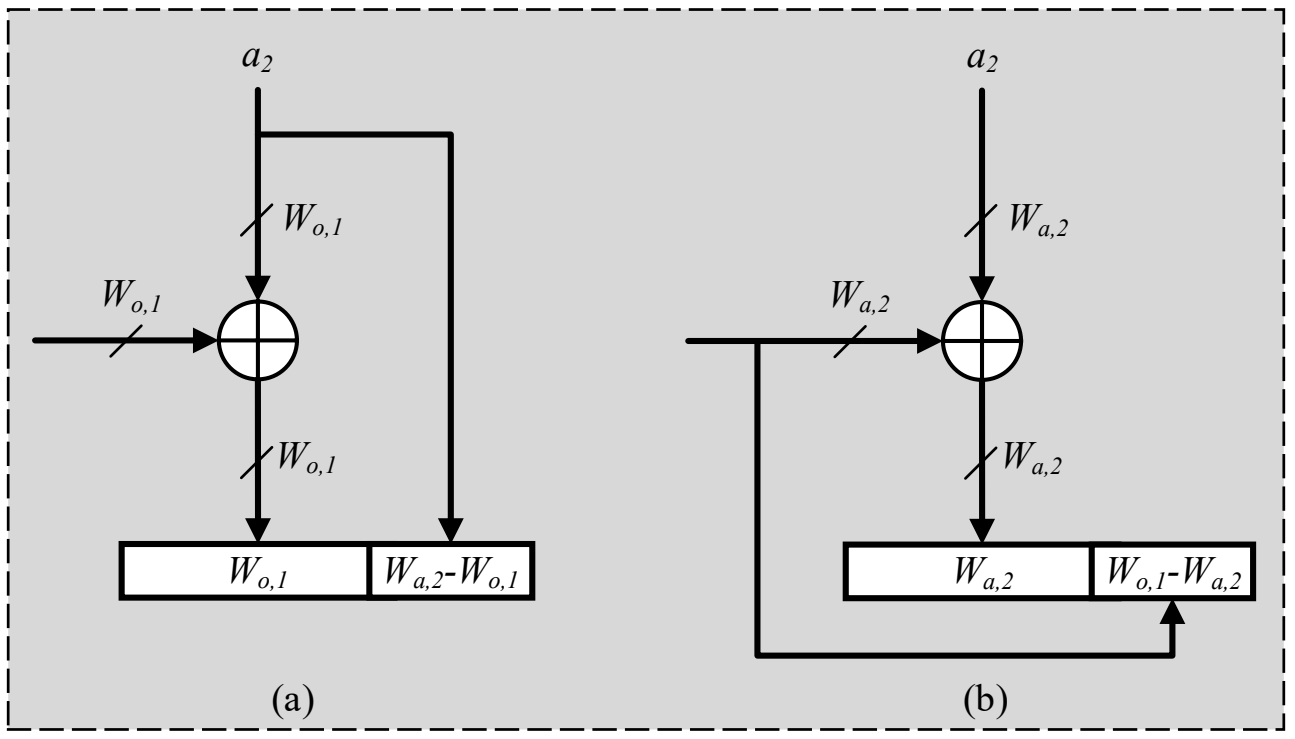

Fig. 3 Concatenation-based adder structure for resource reduction, illustrating two methods of FWL reduction.

After determining the FWLs of the coefficients, each coefficient must be quantized accordingly. Traditional approaches first compute high-precision, unquantized coefficients in software, followed by quantization using common methods such as round, floor, or ceil. However, when the multiplier output has a small FWL, the computed results may significantly deviate from the best values, leading to a noticeable increase in $MAE_{hard}$. In such cases, the optimal quantized coefficient may exhibit a significant deviation, often extending far beyond the quantization space defined by floor and ceil operations. Although the fine-tuning strategy adopted by QPA outperforms previous methods, it does not establish a proper search range for optimal quantization. For tasks with high precision requirements, this limitation may lead to a substantial increase in the number of segments—or even failure to meet accuracy constraints under fixed FWL settings.

Generally, for tasks with low computation precision requirements, directly applying round or floor quantization to the floating-point parameters obtained from the software is often sufficient. However, for tasks demanding high computation precision, a significant deviation may exist between the calculated floating-point coefficients and their optimal quantized values. Under this circumstance, dedicating extensive time to searching for high-precision coefficients prior to quantization can be redundant or inefficient. In the following, we introduce our proposed quantization strategy.

We begin by analyzing the truncation error introduced by limiting the output FWL of the multiplier. Consider the first-stage multiplier as an example. Since $W_{o,1} \le W_{a,1} + W_i$, any fractional bits beyond the $W_{o,1}$ will be directly truncated. Due to the constantly changing input values, direct analysis is challenging. We therefore focus on optimizing the truncation error by analyzing the multiplication coefficients. From the multiplication process, it can be observed that the truncated portion of the result is primarily generated by the multiplication of the lower $W_{a,1} + W_i - W_{o,1}$ bits of the coefficient a with the input $x_q$. Thus, the truncation error introduced by the multiplier only affects this portion of the computation.

Taking both coefficient quantization error and multiplier truncation error into account, the effective offset range of the coefficient extends to $W_{a,1} + W_i - W_{o,1}$ bits. Any value within this range could potentially yield the optimal result. Therefore, prior to quantization, only the integer part and the upper $W_{o,1} - W_i$ bits of the fractional part of the coefficient need to be determined, while the lower $W_{a,1} + W_i - W_{o,1}$ bits of the fractional part are initially set to 0. The ultimate result is then found through an exploration of the full quantization space.

Based on this strategy, the coefficient quantization formula in FQA is defined as:

$$a_{1,q} \in \{\tilde{a}_{1,q} \mid \tilde{a}_{1,q} = a_1 + d / 2^{W_{a,1}}, d \in [0, 2^{W_{a,1}+W_i-W_{o,1}}] \cap Z\} \quad (4)$$

where $a_1$ denotes the precomputed coefficient and $d$ represents the offset. Similarly, the quantization formula for the coefficients in subsequent stages is given by:

$$a_{i,q} \in \{\tilde{a}_{i,q} \mid \tilde{a}_{i,q} = a_i + d / 2^{W_{a,i}}, d \in [0, 2^{W_{a,i}+W_{a,i-1}-W_{o,i}}] \cap Z\} \quad (5)$$

It is important to note that truncation errors may cause different quantized values of a coefficient ai to produce the same output. Therefore, multiple coefficient values satisfying the $MAE_{hard}$ condition can be found. By extending the search range of $d$ to $[-2^{W_{a,i}+W_{a,i-1}-W_{o,i}}, 2^{W_{a,i}+W_{a,i-1}-W_{o,i}+1}]$, more candidate values of a that yield the same output can be identified and stored only once, thereby reducing memory redundancy at the cost of increased software runtime.

In fact, a polynomial function may correspond to multiple optimal fitting solutions, meaning that the quantized coefficients produced by FQA can include several valid values. It is worth noting that in certain segments, the $MAE_{hard}$ may be smaller than the target maximum absolute error ($MAE_t$), which implies a broader feasible range for the coefficients. FQA can accurately identify the valid coefficient range that satisfies the approximation requirement. For different segments with overlapping coefficient ranges, the coefficients can be unified and stored only once, thereby reducing hardware overhead through this optimization strategy.

The coefficients $b$ is to adjust the output of the final-stage multiplier, ensuring that the fitted function approximates the target nonlinear function in the direction determined by the rounding operation.

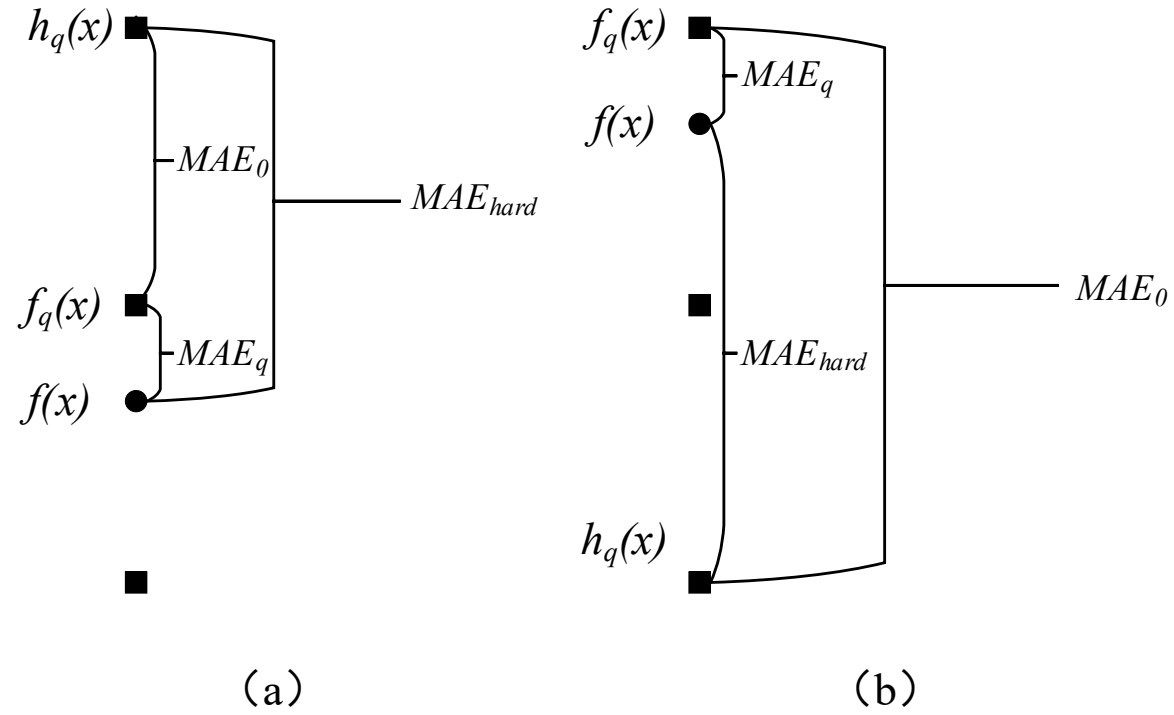

Fig. 4 Illustration of the PPA approximation error and the quantization bias.

Initially, $b$ can be set to 0 and is determined after all other coefficients have been obtained, using an error-flattened strategy followed by rounding, without requiring any further adjustments.

Considering that the target function inherently incurs error during the quantization process, the maximum error within the interval is referred to as the maximum quantization error ($MAE_q$). $MAE_q$ is independent of the PPA procedure and unavoidable. Once the output FWL is determined, the corresponding $MAE_q$ can be calculated using the following expression:

$$MAE_q = \max(| f_q(x) - f(x) |) \tag{6}$$

Since $MAE_{hard}$ includes the quantization error $MAE_q$, it does not directly reflect the approximation accuracy of the PPA method. To address this, we define a direct approximation error for PPA, denoted as $MAE_0$, given by:

$$MAE_0 = \max(| f_q(x) - h_q(x) |) \tag{7}$$

where $f_q(x)$ is quantized using the round method. Clearly, the output FWL determines the upper limit of PPA accuracy. The theoretical lower bound of the PPA approximation error is denoted as $MAE_q$, which satisfies $MAE_q \le MAE_{hard}$, with equality when $MAE_0 = 0$.

Typically, $MAE_{hard} = MAE_0 + MAE_q$; however, due to the inability of methods like QPA and PLAC to determine optimal multiplier coefficients, they may fail to guide the polynomial output sufficiently close to the quantized value $f_q(x)$, instead biasing toward suboptimal quantization directions. As a result, the actual $MAE_0$ obtained by these methods may exceed $MAE_{hard}$ under certain conditions, at which time $MAE_{hard} = MAE_0 - MAE_q$, as shown in Fig. 4(b).

In contrast, FQA can determine the optimal quantized coefficient range under a given FWL, ensuring $MAE_0 = 0$. Therefore, FQA consistently achieves lower approximation error than prior works under the same WL constraints. Due to the enlarged design space explored by FQA, its software runtime is generally longer than that of prior works. However, this additional computation time is typically acceptable and justified, especially considering that the offset introduced by multiplier truncation error is often non-negligible.

### *B. TBW Method*

The bisection method adopted in PLAC computes the MAE across the entire input interval during each initial segmentation step. However, the actual segmentation result often occupies only a fraction of the interval, leading to substantial redundant computation. Moreover, as the FWL of $x$ increases, memory consumption for MAE evaluation grows significantly, posing challenges to the segmentation. Given that, in typical scenarios, the results yielded by non-uniform segmentation and uniform segmentation are generally comparable for an equivalent number of partitions, this presents a potential optimization space for the bisection method. We introduce TBW, which significantly reduces computational cost and memory usage by leveraging a pre-defined target segment number. TBW requires estimate a reasonable range for the target segment count. For FQA, the result obtained by setting $d$ to 0 can be used as a reference value, and the final segment count is typically less than this reference value. For pre-designed configurable hardware, the target segment count can be directly set to the maximum segment count supported by the hardware to fully utilize resources. Notably, although TBW is developed in the context of FQA, it can also be applied to other PPA methods. The process of TBW method is illustrated in Fig. 5. It consists of two components: an inner loop and an outer loop. The inner loop performs the optimization of a single segment, while the outer loop manages the iteration across the entire domain. In this context:

$i$ denotes the current segment index, and $j$ indicates the starting point of the remaining interval to be segmented.

*tSEG* represents the estimated target number of segments.

*INT* is the interval width based on uniform segmentation with *tSEG* segments.

*NUM* is the total length of the interval.

*sp* and *ep* denote the start point and end point of the current segment, respectively.

*lp* and *rp* define the left and right boundaries of the search window.

*rflag* is a flag indicating whether the current search window should be expanded.

The execution flow of the FQA method using the TBW is as follows:

1. Determine *tSEG*. For FQA, the maximum segment count, $SEG_{max}$, under the current $MAE_t$ can be determined by setting $d$ to 0. The final segmentation result is often less than this upper bound. Typically, to facilitate subsequent calculations, *tSEG* is set to an integer power of 2 within the target segment number range [0, $SEG_{max}$]. Initialize $i$ to 0, $j$ to 1, *rflag* to 1, and *ep* to 0.

2. Set *lp* of the search window to $j$, and *rp* of the search window to *NUM*. Set *sp* to $j$. Determine if *ep* is less than *NUM* - *INT*. If the condition is met, increment *ep* by *INT*; otherwise, set *ep* to the midpoint of (*lp* : *rp*).

3. Compute $MAE_{hard}$ of the polynomial within the interval *x(sp : ep)* using the FQA-based algorithm, and check if the $MAE_t$ requirement is satisfied.

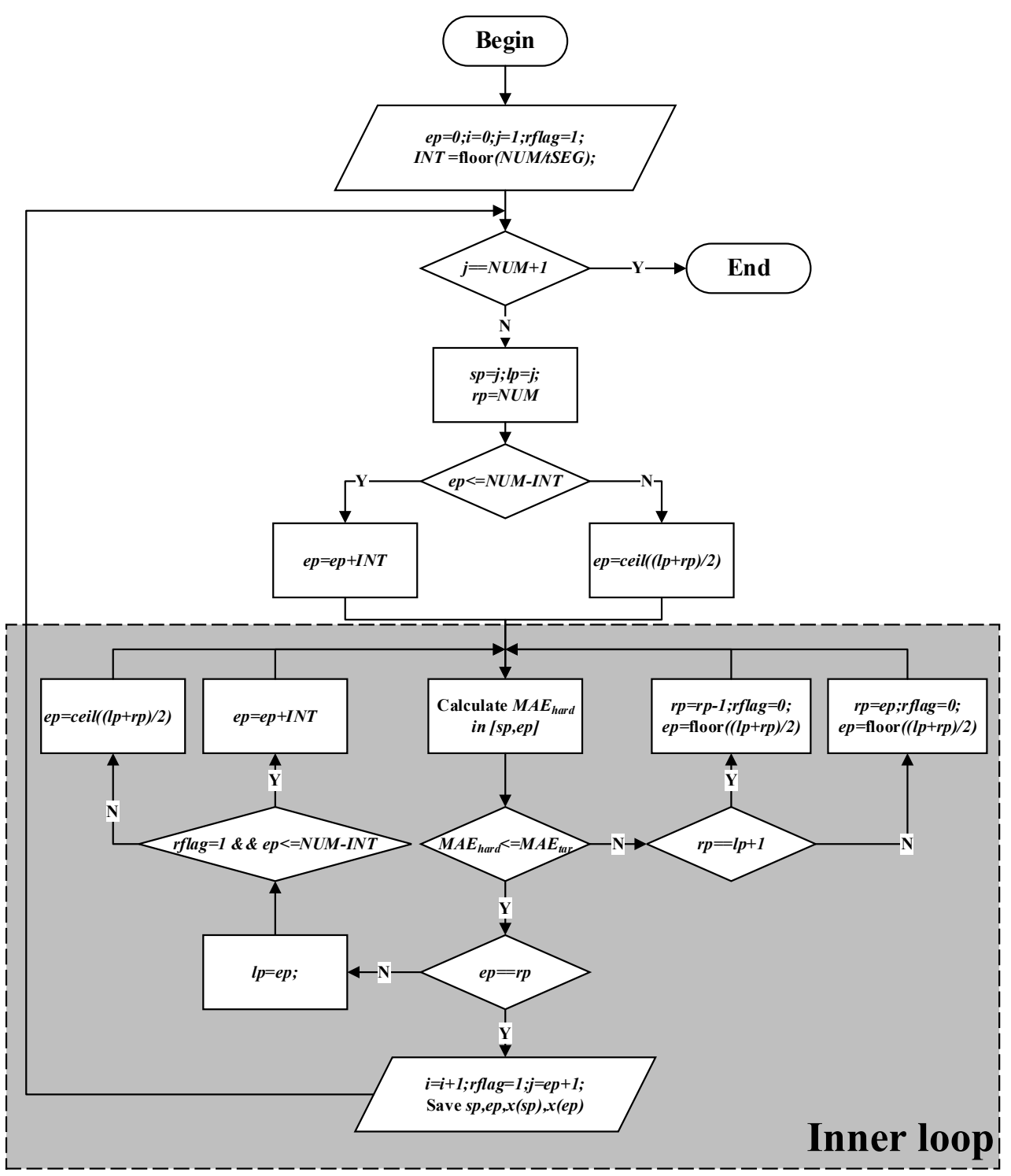

Fig. 5 Procedure of the TBW

Condition A: If the $MAE_t$ requirement is satisfied, check the maximum width condition *ep == rp*. If the maximum width condition holds, the inner loop terminates. Otherwise, the system enters the Segment Interval Expansion Process. In this process, *lp* is first updated to ep. Then, check if *rflag* == 1 and *ep<=NUM-INT* are simultaneously satisfied. If the condition is met, *ep* is incremented by INT; otherwise, ep is shifted to the right to the midpoint of (*lp : rp*), which allows for a more precise determination of the optimal *ep* and reduces computation.

Condition B: If $MAE_{hard}$ does not meet the $MAE_t$ requirement, the system enters the Segment Interval Shrinkage Process. First, check if *rp==lp*+1 is true. If true, *rp* is decremented by 1, *rflag* is set to 0, and *ep* is shifted left to the midpoint of (*lp : rp*). If not true, *rp* is updated to *ep*. The update values for *rflag* and *ep* are not affected by this conditional check. This step addresses the extreme case where the segment interval contains only one point.

4. Repeat Step 3. When the maximum width condition is met, the inner loop terminates. Update and save *i* as *i* + 1, and store key information such as the approximation coefficient $a_n$, b, *sp*, and ep. Finally, update *j* to *ep* + 1 and *rflag* to 1.

5. Repeat the segmentation process in Step 4 until *j* equals *NUM* + 1.

It is important to note that the bisection method used in PLAC cannot handle extreme cases where a segment contains only a single point, which may prevent achieving $MAE_0$ =0. In scenarios with high precision requirements, it is possible for certain intervals to shrink to a single point. However, this does not imply that the total number of segments becomes unacceptable. Especially for pre-manufactured reconfigurable hardware, the hardware resources should be utilized as much as possible to reduce the calculated $MAE_{hard}$, provided that the maximum number of hardware segments is not exceeded.

The TBW method significantly reduces the initial computation overhead of the original bisection-based segmentation algorithm. Taking the segmentation of the first interval as an example, under the same total segment count, the end point of the first non-uniform segment, denoted as $p_{uneven1}$, typically falls either to the left or right of the first end point of the uniform segmentation, denoted as $p_{even1}$. In most cases, $p_{uneven1}$ does not exceed the end point of the second uniform segment, $p_{even2}$. For computational convenience, we set $tSEG = 2^N$. When $p_{uneven1}$ lies to the left of $p_{even1}$, the TBW method can determine the right boundary of the segment in just one computation, while the PLAC method requires up to *N* iterations to reach the same conclusion. The time ratio in this case is:

$$2^{N+1} - 1 \tag{8}$$

After determining the right boundary of peven1, the remaining computation procedures for both TBW and PLAC are essentially the same. The total time ratio for computing the first segment becomes:

$$1 + \frac{2^{N+1}\text{-}2}{W_i - N + 2^{N-W_i}} \tag{9}$$

When $p_{uneven1}$ lies to the right of $p_{even1}$, the TBW method requires only two computations to determine the right boundary, while PLAC still requires N iterations. The total time ratio for computing the first segment in this scenario becomes:

$$1 + \frac{2^{N+1} - 4}{W_i - N + 2 + 2^{N-W_i}} \tag{10}$$

From the derived expressions, the larger the value of N, the greater the speedup achieved by TBW. For instance, when $W_i$ = 8 and *N* = 4, the time ratio for determining the right boundary of $p_{even1}$ equals 31, resulting in a speedup factor of 8.4 in the case where $p_{uneven1}$ lies to the right of $p_{even1}$. Similarly, if $p_{uneven1}$ lies to the left, a speedup of 5.6 is obtained.

### *C. FQA-On Scheme*

In this section, we propose the FQA-On scheme based on mainstream PPA computing circuits, where *n* represents the order of the polynomial. The computational circuit for FQA-On typically comprises *n* multipliers and *n* adders, as illustrated in Fig. 2. Typically, *n* is configured as 1, making FQA-O1 a type of PWL method. When higher computational precision is required, using FQA-O1 can lead to a sharp increase in the number of segments. In such cases, *n* is configured as 2 to achieve a more efficient segmentation scheme. In practice, configurations with *n* > 2 are rare because multipliers are often the most area-intensive components in PPA-optimized designs, and adding more multipliers usually results in excessive hardware area overhead.

Algorithm 1 describes the computation process of FQA-On. The fitting algorithm used to determine the pre-quantization coefficients is not restricted. In this work, the Remez algorithm is chosen to obtain the initial coefficient values. Taking FQA-O1 as an example, since the FQA quantization method only requires the  fractional bit of the coefficient to be accurate, the number of iterations in the Remez algorithm can be reduced, thus saving pre-quantization computation effort.

To demonstrate the optimal coefficient offset caused by quantization and multiplier truncation errors, we select the Sigmoid function as the fitting target function over the interval [0,1). We first find the optimal pre-quantization coefficient values. Then, we set the FWLs as $W_i$=8, $W_{a,1}$=8, $W_b$=8, and $W_{o,1}$=8, and complete the final quantization and segmentation using FQA-O1. The segmentation results are presented in TABLE I. In segment 9, the coefficients yielding the $MAE_t$ exhibit a deviation ranging from a minimum of 69 to a maximum of 131 relative to the optimal pre-quantization values. This finding indicates that relying solely on conventional quantization methods like rounding or simple ± 1 fine-tuning cannot achieve the optimal quantized coefficient values.

We conducted a detailed comparison between FQA-O1, FQA-O2, and other existing works, with the results summarized in TABLE II and TABLE III. QPA and PLAC employ segmentation algorithms that may not account for the degenerate case where a segment interval contains only a single point, potentially failing to reach the theoretical precision limit in some cases. To enable a fairer comparison of segmentation results, we replaced the segmentation algorithms of QPA and PLAC with the TBW to determine the minimum number of segments required under the precision limit. Our proposed algorithm significantly reduces the number of segments, achieving superior segmentation results across different output precisions. Furthermore, in the scenario with an input FWL of 8-bit and an output FWL of 16-bit, even though FQA-O1 shows considerable improvement over other methods, the number of segments remains relatively high. This suggests that the FQA-O2 scheme is more suitable for this case.

The determination of FWLs represents another critical challenge in the FQA. Selecting appropriate FWLs is essential for minimizing hardware area and power consumption. In practice, the multiplier FWL significantly influences the required number of segments; an inadequate multiplier FWL may lead to a prohibitive increase in the segment count, offsetting the initial hardware savings. In contrast, the adder FWL exerts a relatively minor impact on segmentation. Given that multipliers incur substantially higher hardware overhead than adders, we adopt an FWL optimization strategy that prioritizes the area reduction of multipliers followed by adders. The design flow for the FWLs of FQA-On is detailed below:

Step 1: Initialization.

Determine $W_i$ and $W_{o,final}$ based on task requirements. Set the FWLs of all other parts to relatively large values. Typically, $W_b$ and $W_{o,n}$ are set equal to $W_{o,final}$, and other FWLs are set equal to $W_i$.

Step 2: Multiplier FWL Optimization.

Determine the FWLs of the multiplier stages sequentially,

**Algorithm** 1 FQA-On Algorithm

Input: pre-quantization coefficients $a_1, a_2, \ldots, a_n, b$

Output: $MAE_{hard}$, the optimal range of quantization coefficients $a_{1,q}, a_{2,q}, \ldots, a_{n,q}, b_q$

1. $\tilde{a}_{i,q} = a_i + d / 2^{W_{a,i}}, d \in [0, 2^{W_{a,i}+W_{a,i-1}-W_{o,i}}] \cap Z$
2. $h = \tilde{a}_{1,q}$
3. **for** $i$ = 1: $n$-1 **do**
4. $\quad h = (h \times x_q \times 2^{W_{o,i}}) / 2^{W_{o,i}} + \tilde{a}_{i+1,q}$
5. **end for**
6. $h = (h \times x_q \times 2^{W_{o,n}}) / 2^{W_{o,n}}$
7. $E_0 = f(x_q) - h$
8. $b = (\max(E_0) + \min(E_0)) / 2$
9. $b_q = round(b \times 2^{W_b}) / 2^{W_b}$
10. $E_f = f(x_q) - h - b_q$
11. $(MAE_{hard}, d_{best}) = \min(\max(abs(E_f)))$
12. $a_{i,q} \in \{\tilde{a}_{1,q} \mid \tilde{a}_{1,q} = a_i + d / 2^{W_{a,i}}, d \in d_{best}\}$

following the order from Mn to M1. The first step is to determine $W_{o,n}$. Decrease $W_{o,n}$ while keeping other FWLs constant. Calculate the current coefficient Look-Up Table (LUT) using the FQA algorithm. Fix $W_{o,n}$ when the LUT size starts to increase. Next, determine $W_{m,n}$. Decrease $W_{m,n}$. Simultaneously, reduce all FWLs preceding $W_{m,n}$ to be equal to $W_{m,n}$. Calculate the current LUT. Fix $W_{m,n}$ when the LUT size starts to increase. Repeat Step 2 until all relevant FWLs for the multipliers ($W_{a,1}$, $W_{o,1}$ to $W_{o,n}$, $W_{m,2}$ to $W_{m,n}$) are fixed.

Step 3: Adder FWL Optimization.

Determine the FWLs of the adder stages sequentially, following the order from A1 to An. First, compare $W_{a,2}$ and $W_{o,1}$. If $W_{a,2} > W_{o,1}$, the A1 FWL is set to $W_{o,1}$. If $W_{a,2} = W_{o,1}$, decrease $W_{a,2}$. The FWL of A1 is set to $W_{a,2}$. Calculate the LUT using the FQA algorithm. Fix $W_{a,2}$ when the LUT size starts to increase. Repeat Step 3 until all relevant FWLs for the adders ($W_{a,2}$ to $W_{a,n}$, $W_b$) are fixed.

This methodology can determine a near-optimal hardware design. However, the drawback is that as $n$ increases, the design space becomes significantly larger. In fact, due to the large area occupied by multiple multipliers, configurations involving more than two multipliers are rarely used in practical applications.

*D. FQA-Sm-On Scheme*

For specific scenarios where the required $MAE_{hard}$ is not strictly demanding, it is feasible to replace the first-stage multiplier with a shifter-adder network to reduce the hardware area. Previous architectures, such as ML-PLAC, limit the WL of the multiplier coefficient when designing the shifter-adder, making the number of shifters equal to the coefficient WL. Furthermore, they do not adequately address the coefficient offset issue during the quantization stage, which leads to lower computational precision. QPA only slightly alleviates the quantization error through fine-tuning but does not fully

TABLE I

THE RESULTS OF FITTING THE SIGMOID FUNCTION USING FQA-O1 IN THE [0,1) INTERVAL

| No. of Seg | 1 | 2 | 3 | 4 | 5 | 6 | 7 | 8 | 9 | 10 | 11 | 12 | 13 | 14 | 15 | 16 | 17 | 18 |
|---|---|---|---|---|---|---|---|---|---|---|---|---|---|---|---|---|---|---|
| $a_1$ ($\times 10^{-3}$) | 336 | 289 | 270 | 262 | 238 | 242 | 250 | 258 | 504 | 219 | 227 | 242 | 137 | 176 | 188 | 211 | 223 | 254 |
| $b$ ($\times 10^{-3}$) | 500 | 500 | 500 | 500 | 504 | 504 | 500 | 496 | 371 | 516 | 512 | 500 | 578 | 551 | 543 | 523 | 512 | 480 |
| $x_q(xs)$ ($\times 10^{-3}$) | 0 | 23 | 55 | 102 | 180 | 262 | 371 | 469 | 500 | 512 | 570 | 707 | 742 | 762 | 816 | 891 | 945 | 984 |
| $x_q(xe)$ ($\times 10^{-3}$) | 20 | 51 | 98 | 176 | 258 | 367 | 465 | 496 | 508 | 566 | 703 | 738 | 758 | 813 | 887 | 941 | 980 | 996 |
| $MAE_{hard\text{-}min}$ ($\times 10^{-4}$) | 20 | 20 | 19 | 19 | 18 | 19 | 19 | 18 | 16 | 19 | 19 | 18 | 17 | 19 | 18 | 19 | 17 | 17 |
| $Dev_{a1,\min}$ | 23 | 11 | 6 | 4 | -2 | 0 | 3 | 6 | 69 | -3 | 1 | 6 | -20 | -10 | -5 | 2 | 6 | 15 |
| $Dev_{a1,\max}$ | 39 | 13 | 6 | 4 | -2 | 3 | 3 | 12 | 131 | -3 | 1 | 13 | 8 | -5 | -5 | 2 | 8 | 35 |

TABLE II

COMPARISON OF PIECEWISE LINEAR APPROXIMATION METHODS

| Function | Method | Target Interval | $W_i$ | $W_{a,1}$ | $W_{o,1}$ | $W_b$ | $W_{o,final}$ | $MAE_{hard}$ | No. of Seg | Improvement of Seg |
|---|---|---|---|---|---|---|---|---|---|---|
| sigmoid(x) | FQA-O1 | [0,1) | 8 | 7 | 8 | 8 | 8 | 1.953*e-3 | 18 | - |
| | QPA-G1 | | | 8 | 8 | 8 | | | 60 | -70.0% |
| | PLAC | | | 8 | 8 | 8 | | | 144 | -87.5% |
| | FQA-O1 | | | 16 | 16 | 14 | 16 | 7.599*e-6 | 33 | - |
| | QPA-G1 | | | 16 | 16 | 16 | | | 45 | -26.7% |
| tanh(x) | FQA-O1 | [0,1) | 8 | 8 | 8 | 8 | 8 | 1.945*e-3 | 15 | - |
| | QPA-G1 | | | 8 | 8 | 8 | | | 34 | -55.9% |
| | PLAC | | | 8 | 8 | 8 | | | 98 | -84.7% |
| | FQA-O1 | | | 14 | 16 | 16 | 16 | 7.606*e-6 | 79 | - |
| | QPA-G1 | | | 16 | 16 | 16 | | | 86 | -8.1% |

TABLE III

COMPARISON OF PIECEWISE QUADRATIC APPROXIMATION METHODS

| Function | Method | Target Interval | $W_i$ | $W_{a,1}$ | $W_{o,1}$ | $W_{a,2}$ | $W_{o,2}$ | $W_b$ | $W_{o,final}$ | $MAE_{hard}$ | No. of Seg | Improvement of Seg |
|---|---|---|---|---|---|---|---|---|---|---|---|---|
| sigmoid(x) | FQA-O2 | [0,1) | 8 | 6 | 8 | 8 | 8 | 8 | 8 | 1.953*e-3 | 10 | - |
| | QPA-G2 | | | 8 | 8 | 8 | 8 | 8 | | | 60 | -83.3% |
| | FQA-O2 | | | 8 | 16 | 16 | 16 | 16 | 16 | 7.599*e-6 | 12 | - |
| | QPA-G2 | | | 8 | 16 | 16 | 16 | 16 | | | 23 | -47.8% |
| tanh(x) | FQA-O2 | [0,1) | 8 | 8 | 8 | 6 | 8 | 8 | 8 | 1.945*e-3 | 8 | - |
| | QPA-G2 | | | 8 | 8 | 8 | 8 | 8 | | | 10 | -20.0% |
| | FQA-O2 | | | 8 | 16 | 16 | 16 | 16 | 16 | 7.606*e-6 | 16 | - |
| | QPA-G2 | | | 8 | 16 | 16 | 16 | 16 | | | 30 | -46.7% |

explore the optimization space. Based on the strategy of replacing the multiplier with a shifter-adder network, we propose FQA-Sm-On strategy, where $m$ represents the number of shifters and $n$ denotes the polynomial order. In FQA-Sm-On, the first-stage multiplier is replaced by $m$ shifters and $m$-1 adders, while the subsequent stages (from the second multiplier onwards) remain identical to FQA-On. The structure is illustrated in Fig. 6. Unlike the ML-PLAC scheme, FQA-Sm-On does not restrict the multiplier coefficient WL to be within the number of shifters $m$. Instead, it only requires that the hamming weight of the coefficient does not exceed $m$.

In this context, the first-stage quantization formula for FQA is modified as follows:

$$a_{1,q} \in \{\tilde{a}_{1,q} \mid \tilde{a}_{1,q} = a_1 + d / 2^{W_{a,1}}, d \in [0, 2^{W_{a,1}+W_i-W_{o,1}}] \cap Z, \; w_H(\tilde{a}_{1,q}) \le m\} \tag{11}$$

where $w_H(\tilde{a}_{1,q})$ is the hamming weight of $\tilde{a}_{1,q}$, and the other coefficients remain unchanged. Due to the potentially large Hamming weight of the first-stage output, it is generally difficult to replace the second-stage and subsequent multipliers with shifter-adder networks.

The FQA-Sm-On algorithm is described in Algorithm 2. Since the number of shifters $m$ constrains the exploration range of the quantized coefficient values, the computational complexity of FQA-Sm-On is usually lower than that of FQA-On. Furthermore, FQA-Sm-On employs the same FWL optimization strategy as FQA-On.

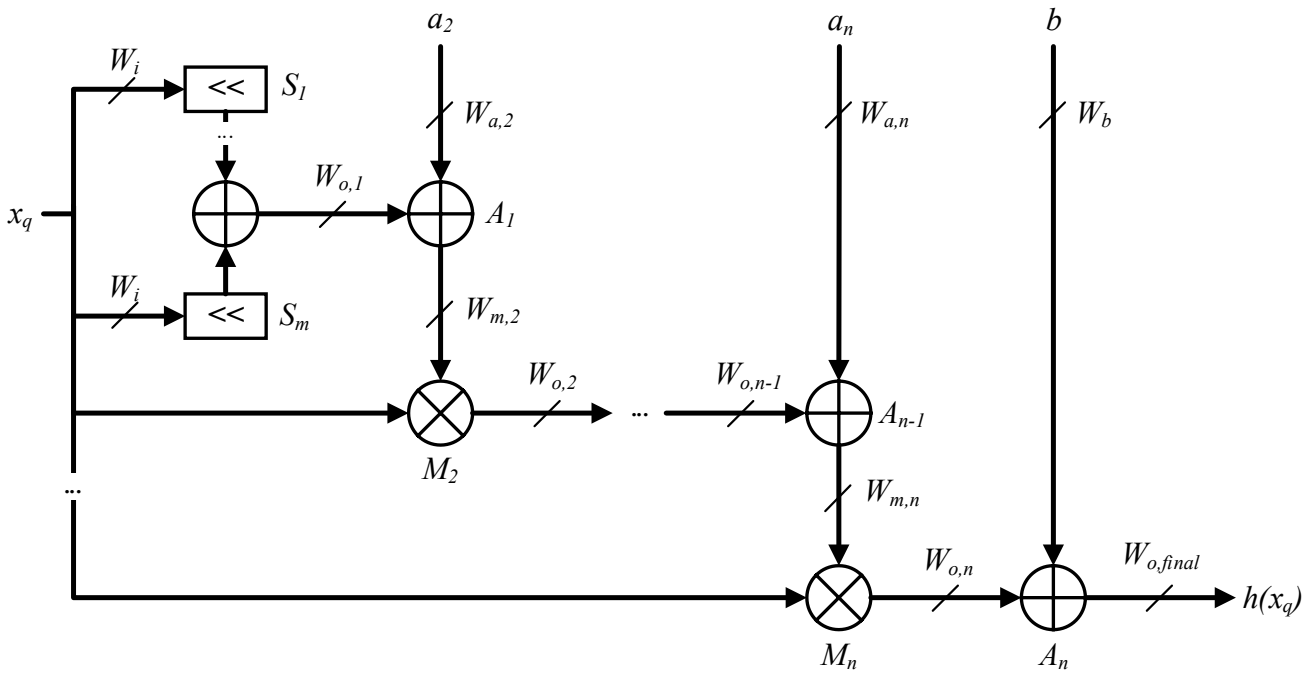

Fig. 6 Calculation circuit corresponding to FQA-Sm-On strategy

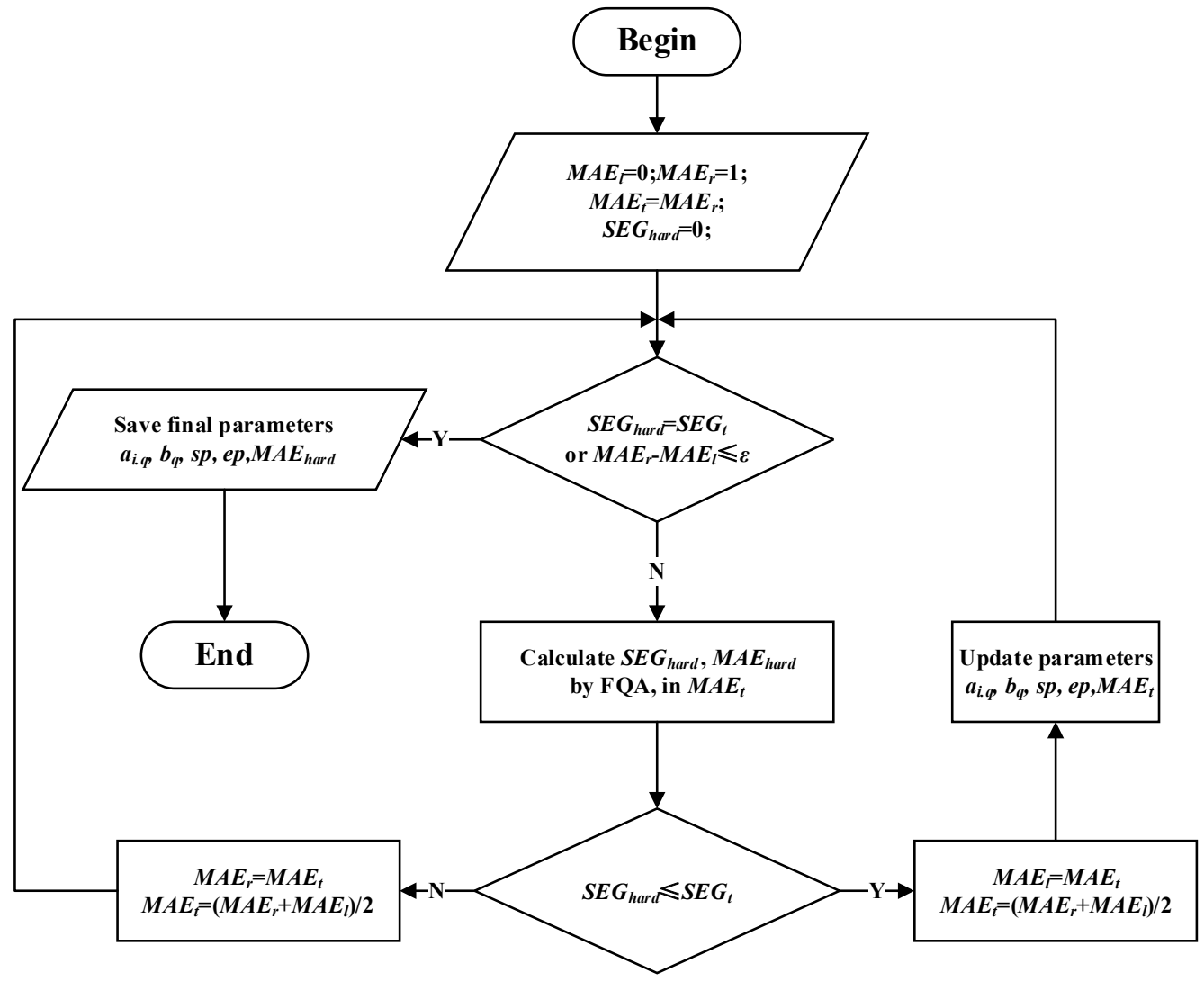

Fig. 7 The proposed hardware-constrained PPA workflow.

FQA-Sm-On thoroughly explores the optimization space, and its quantization process encompasses the optimal coefficient range. Consequently, it can significantly reduce the computational $MAE_{hard}$—and thus the number of segments, while keeping the number of shifters constant. Although the indexing part of the shifters may be more complex in certain situations, the advantage gained from the reduced number of segments outweighs the negative impact of the increased complexity of the indexer.

We compared the segmentation performance of FQA-Sm-O1 with QPA-M1 and ML-PLAC, and the results are shown in TABLE IV. FQA-Sm-O1 demonstrates a clear advantage in segmentation results. Under certain conditions, FQA-Sm-O1 achieves the same number of segments as FQA-O1, while certainly incurring less hardware overhead than FQA-O1. We also tested the implementation of FQA-Sm-O2, with results presented in TABLE V. FQA-Sm-O2 also achieves the same number of segments as FQA-O2 in some cases, providing a noticeable optimization in hardware area.

### *E. Hardware-Constrained PPA Workflow*

To accommodate various NAFs, a reconfigurable hardware design offers significantly broader applicability.

**Algorithm 2** FQA-Sm-On Algorithm

Input: pre-quantization coefficients $a_1, a_2, \ldots, a_n, b$

Output: $MAE_{hard}$, the optimal range of quantization coefficients $a_{1,q}, a_{2,q}, \ldots, a_{n,q}, b_q$

1. $\tilde{a}_{1,q} = a_1 + d / 2^{W_{a,1}}, d \in [0, 2^{W_{a,1}+W_i-W_{o,1}}] \cap Z, w_H(\tilde{a}_{1,q}) \le m$
2. $\tilde{a}_{i,q} = a_i + d / 2^{W_{a,i}}, d \in [0, 2^{W_{a,i}+W_{a,i-1}-W_{o,i}}] \cap Z, i \ge 2$
3. $h = \tilde{a}_{1,q}$
4. **for** $i = 1 : n\text{-}1$ **do**
5. $\quad h = (h \times x_q \times 2^{W_{o,i}}) / 2^{W_{o,i}} + \tilde{a}_{i+1,q}$
6. **end for**
7. $h = (h \times x_q \times 2^{W_{o,n}}) / 2^{W_{o,n}}$
8. $E_0 = f(x_q) - h$
9. $b = (\max(E_0) + \min(E_0)) / 2$
10. $b_q = round(b \times 2^{W_b}) / 2^{W_b}$
11. $E_f = f(x_q) - h - b_q$
12. $(MAE_{hard}, d_{best}) = \min(\max(abs(E_f)))$
13. $a_{i,q} \in \{\tilde{a}_{1,q} \mid \tilde{a}_{1,q} = a_i + d / 2^{W_{a,i}}, d \in d_{best}\}$

For post-fabrication reconfigurable logic, where the hardware resources are already silicon-defined, the primary challenge lies in maximizing computational precision (i.e., minimizing $MAE_{hard}$) within the confines of fixed hardware constraints. The preceding analysis confirms that FQA effectively explores the coefficient space to reach an optimal trade-off: it yields the lowest $MAE_{hard}$ under a specific segment constraint or, conversely, minimizes the required segments for a given $MAE_t$. Therefore, FQA is well-suited for improving the approximation accuracy of resource-constrained or fixed-architecture neural accelerators.

To leverage this capability, we design a hardware-constrained PPA workflow, as illustrated in the Fig. 7. In this flow, $\varepsilon$ denotes the tolerance for the search width, while $SEG_{hard}$ and $SEG_t$ represent the current segment count and the maximum hardware-supported segment, respectively. Since FQA attains the optimal $MAE_{hard}$ for any given $SEG_{hard}$, the design process terminates once $SEG_{hard} = SEG_t$, at which point the $MAE_{hard}$ and the corresponding coefficients are stored.

If the PPA workflow were based on a predefined $MAE_t$, it might lead to two suboptimal scenarios: The resulting number of segments is greater than the number supported by the current hardware configuration, making it impossible to map onto the physical hardware. The resulting number of segments is far less than the number supported by the current hardware, leading to wasted hardware resources, as increasing the number of segments could potentially further reduce $MAE_{hard}$.

Although the proposed workflow is not restricted to the FQA, FQA typically delivers superior performance over existing works by precisely identifying the optimal coefficients for any given segmentation.

TABLE IV

COMPARISON OF MULTIPLIERLESS PIECEWISE LINEAR APPROXIMATION METHODS

| Function | Method | Target Interval | $W_i$ | $W_{a,1}$ | $W_{o,1}$ | $W_b$ | $W_{o,final}$ | $MAE_{hard}$ | No. of Seg | Improvement of Seg |
|---|---|---|---|---|---|---|---|---|---|---|
| sigmoid(x) | FQA-S2-O1 | [0,1) | 8 | 8 | 8 | 8 | 8 | 1.953*e-4 | 24 | - |
| | FQA-S4-O1 | | | 8 | 8 | 8 | | | 18 | 33.3% |
| | QPA-M1 | | | 1 | 8 | 8 | | | 60 | -60.0% |
| | ML-PLAC | | | 1 | 8 | 8 | | | 60 | -60.0% |
| tanh(x) | FQA-S2-O1 | [0,1) | 8 | 7 | 8 | 8 | 8 | 1.945*e-4 | 28 | - |
| | FQA-S4-O1 | | | 8 | 8 | 8 | | | 17 | 64.7% |
| | QPA-M1 | | | 1 | 8 | 8 | | | 52 | -46.2% |
| | ML-PLAC | | | 1 | 8 | 8 | | | 54 | -48.1% |

TABLE V

THE SEGMENTED RESULTS OF FQA-Sm-O2 METHODS

| Function | Method | Target Interval | $W_i$ | $W_{a,1}$ | $W_{o,1}$ | $W_{a,2}$ | $W_{o,2}$ | $W_b$ | $W_{o,final}$ | $MAE_{hard}$ | No. of Seg |
|---|---|---|---|---|---|---|---|---|---|---|---|
| sigmoid(x) | FQA-S3-O2 | [0,1) | 8 | 8 | 8 | 8 | 8 | 8 | 8 | 1.953*e-3 | 10 |
| | FQA-S3-O2 | | | 8 | 16 | 16 | 16 | 16 | 16 | 7.599*e-6 | 12 |
| tanh(x) | FQA-S4-O2 | [0,1) | 8 | 8 | 8 | 6 | 8 | 8 | 8 | 1.945*e-3 | 8 |
| | FQA-S4-O2 | | | 8 | 16 | 16 | 16 | 16 | 16 | 7.606*e-6 | 17 |

For instance, in scenarios where the $MAE_{hard}$ is already low but the segment count remains below the hardware limit, prior PPA methods may struggle to further reduce the error, even when the number of segments is increased.

Our proposed flow effectively utilizes the available resources of the reconfigurable hardware, ensuring that the computational precision reaches its optimum.

## IV. ASIC IMPLEMENTATION AND COMPARISON

We model the proposed hardware architecture using Verilog HDL and synthesize it using Synopsys Design Compiler (DC) under the 65nm TSMC CMOS process. In this section, our target function is the Sigmoid function, with the target interval of [0,1], and the $MAE_t$ is uniformly set to the minimum achievable value for the current precision. In Section IV-A, we use an 8-bit input and 8-bit output and present the implementation results of our proposed hardware architectures, comparing them with previous works. In Section IV-B, we analyze the scenario involving higher precision, where the output is elevated to 16 bits, and discuss the corresponding implementation results.

### *A. 8-bit Precision Implementation Results*

TABLE VI shows the hardware implementation results for both 8-bit input and output, with the $MAE_{hard}$ of 1.953*e-4. It is evident that the FQA method demonstrates a significant advantage in the number of segments. Furthermore, the wider range of derived coefficients offers greater optimization space for cases where multiple segments share the same coefficient, thus providing a huge area advantage compared to previous works.

The area occupied by FQA-O1 and FQA-O2 is comparable. Typically, FQA-O2 uses two multipliers, which would generally incur larger resource usage for lower precision. However, in the current scenario, although FQA-O1 has already significantly reduced the number of segments compared to QPA-G1, it still requires 18 segments, making its hardware area comparable to, or even larger than, the area of FQA-O2. The FQA-Sm-O1 scheme achieves the same number of segments as FQA-O1, but the multiplier is replaced by 4 shifters and 3 adders, resulting in superior hardware area efficiency. Similarly, the FQA-Sm-O2 scheme achieves the same number of segments as FQA-O2, but the first-stage multiplier is replaced by only 3 shifters and 2 adders. This results in the lowest area overhead overall. However, due to the increased complexity of the computation unit, the power consumption is higher than that of the FQA-O1 and FQA-Sm-O1 schemes.

The experimental results indicate that different FQA strategies have distinct advantages and disadvantages under low-precision conditions, and the final choice of scheme should be determined based on specific application requirements.

### *B. 16-bit Precision Implementation Results*

TABLE VII displays the hardware implementation results for both 16-bit input and output, with the $MAE_{hard}$ of 7.599*e-6. Under the same approximation order, the FQA scheme consistently outperforms other existing works.

It can be observed that, due to the high target precision, even using FQA-S5-O1 still necessitates many segments to satisfy the error requirements. In this high-precision context, the first-order piecewise linear approximation using the shifter-multiplier replacement is no longer feasible. Even the FQA-O1 scheme, which utilizes a full multiplier and increases the coefficient precision to 16 bits, requires 33 segments, leading to substantial hardware area consumption. This suggests that a first-order approximation scheme is no longer appropriate at this precision level. FQA-O2, when targeting a 16-bit output, only increases the number of segments by 2 compared to the 8-bit output result, demonstrating a clear advantage. However, due to the increased WL of the multipliers, the multiplier area overhead increases dramatically, making it sub-optimal.

TABLE VI

ASIC IMPLEMENTATION RESULTS OF 8-BIT OUTPUT PRECISION

| Method | $W_i$ | $W_{a,1}$ | $W_{o,1}$ | $W_{a,2}$ | $W_{o,2}$ | $W_b$ | $W_{o,final}$ | $MAE_{hard}$ | No. of Seg | Area($\mu m^2$) | Delay (ns) | Power (mW) |
|---|---|---|---|---|---|---|---|---|---|---|---|---|
| FQA-O1 | | 7 | 8 | - | - | 8 | | | 18 | 1581.2 | 1.67 | 0.2185 |
| QPA-G1 | | 8 | 8 | - | - | 8 | | | 60 | 4919.2 | 2 | 0.8956 |
| PLAC | | 8 | 8 | - | - | 8 | | | 144 | 11419.6 | 1.98 | 1.7293 |
| FQA-S2-O1 | | 8 | 8 | - | - | 8 | | | 24 | 1595.2 | 1.48 | 0.1777 |
| FQA-S4-O1 | | 8 | 8 | - | - | 8 | | | 18 | 1398.4 | 1.47 | 0.1849 |
| QPA-M1 | 8 | 1 | 8 | - | - | 8 | 8 | 1.953*e-4 | 60 | 3794.8 | 1.8 | 0.6484 |
| ML-PLAC | | 1 | 8 | - | - | 8 | | | 60 | 3794.8 | 1.8 | 0.6484 |
| FQA-O2 | | 6 | 8 | 8 | 8 | 8 | | | 10 | 1496.8 | 1.7 | 0.3012 |
| QPA-G2 | | 8 | 8 | 8 | 8 | 8 | | | 60 | 6247.2 | 2 | 1.103 |
| FQA-S1-O2 | | 8 | 8 | 8 | 8 | 8 | | | 13 | 1360.79 | 1.79 | 0.2247 |
| FQA-S3-O2 | | 8 | 8 | 8 | 8 | 8 | | | 10 | 1294 | 1.62 | 0.26 |

TABLE VII

ASIC IMPLEMENTATION RESULTS OF 16-BIT OUTPUT PRECISION

| Method | $W_i$ | $W_{a,1}$ | $W_{o,1}$ | $W_{a,2}$ | $W_{o,2}$ | $W_b$ | $W_{o,final}$ | $MAE_{hard}$ | No. of Seg | Area($\mu m^2$) | Delay (ns) | Power (mW) |
|---|---|---|---|---|---|---|---|---|---|---|---|---|
| FQA-O1 | | 16 | 16 | - | - | 14 | | | 33 | 4307.59 | 2 | 0.5775 |
| QPA-G1 | | 16 | 16 | - | - | 16 | | | 45 | 5865.6 | 2 | 1.1953 |
| FQA-S5-O1 | | 9 | 16 | - | - | 16 | | | 75 | 6979.6 | 2 | 0.6433 |
| FQA-O2 | 8 | 8 | 16 | 16 | 16 | 16 | 16 | 7.599*e-6 | 12 | 3105.59 | 1.93 | 0.7919 |
| QPA-G2 | | 8 | 16 | 16 | 16 | 16 | | | 23 | 4527.2 | 2 | 1.3405 |
| FQA-S1-O2 | | 8 | 16 | 16 | 16 | 16 | | | 18 | 2989.59 | 2 | 0.5338 |
| FQA-S3-O2 | | 8 | 16 | 16 | 16 | 16 | | | 12 | 2554.4 | 1.98 | 0.5982 |

The FQA-Sm-O2 scheme does not increase the number of segments, while replacing the first-stage multiplier with a shifter-adder network, thereby achieving the best hardware result. This confirms that higher-order approximation methods offer stronger approximation capability, and the optimal range for the high-order coefficients may be wider, enabling the possibility of replacing the multiplier with a shifter-adder network.

## V. Conclusion

In this paper, we propose FQA, which incorporates a high-degree polynomial approximation scheme and a multiplier-less implementation strategy leveraging shifter-adder networks. We further decouple the FWLs of each section and employ a fractional bit concatenation strategy to maximize the optimization of hardware resource overhead. Given that FQA exhibits a significantly larger search space, we also devise TBW to accelerate the computation. Experimental results demonstrate that our proposed scheme achieves the minimum number of segments under a fixed MAE constraint, showing a distinct advantage over existing works such as QPA, PLAC, and ML-PLAC. Furthermore, we present a design flow under fixed hardware resource constraints, which, for hardware implementations of configurable activation functions, maximizes hardware resource utilization while achieving the lowest possible MAE.